\documentclass[prb,twocolumn,showpacs,superscriptaddress,preprintnumbers]{revtex4}
\usepackage{graphicx}
\usepackage{dcolumn}
\usepackage{bm}
\usepackage{wasysym}
\def\momit#1{}

\newcommand{\beq}{\begin{equation}}
\newcommand{\eeq}{\end{equation}}
\newcommand{\beqn}{\begin{eqnarray}}
\newcommand{\eeqn}{\end{eqnarray}}

\begin{document}
\title{Nonequilibrium charge density wave ordering from anomalous velocity \\
in itinerant helical magnets}
\author{Cenke Xu}
\affiliation{Department of Physics, University of California, Berkeley, CA 94720}
\author{J.~E.~Moore}
\affiliation{Department of Physics, University of California, Berkeley, CA 94720} \affiliation{Materials
Sciences Division, Lawrence Berkeley National Laboratory, Berkeley, CA 94720} 
\date{\today}
\begin{abstract}

\momit{We found that in some noncollinear ferromagnetic semiconductors, electrons tend to form charge ordered state
once we add a electric field in certain direction. Whether there is charge ordered state or not can be directly
determined by symmetry argument. Normally, if the local moment has homogenous magnitude, the
homogeneity of charge
is protected by
gauge invariance when there is no electric field, and
certain electric field will break this gauge symmetry.}

The Karplus-Luttinger anomalous velocity is shown to lead to electric-field-induced charge accumulation in
nearly ferromagnetic noncollinear magnets with itinerant electrons, like $MnSi$.  For helical magnetic ordering,
the balance between this accumulation and the Coulomb interaction leads to a nonequilibrium charge density wave
state with the period of the helix, even when such accumulation is forbidden by an approximate gauge-like
symmetry in the absence of electric field.  We compute the strength of such charge accumulation as an example of
how unexpected many-electron physics is induced by the inclusion of the one-electron Karplus-Luttinger term
whenever the local exchange field felt by conduction electrons does not satisfy the current-free Maxwell
equations.

\end{abstract}
\maketitle


Recently the Karplus-Luttinger ``anomalous velocity''  in crystal structures without inversion symmetry has been
the subject of renewed interest.  It has been reinterpreted as an additional term in the semiclassical equations
of motion resulting from a Berry phase in momentum space~\cite{niu2}, and shown to be determined solely by
electronic properties at the Fermi surface~\cite{haldane}, consistent with the spirit of Landau's Fermi liquid
theory.  A natural question is whether the existence of a new term in the single-electron equation of motion
modifies standard interacting electron physics.  We answer this question in the affirmative for long-period
itinerant helimagnets like MnSi and related systems: the anomalous velocity leads to a nonequilibrium
steady-state charge density wave (CDW) transverse to an applied electrical field.

Such ordering is forbidden by symmetry in the absence of an electrical field, but the combination of an applied
spatially constant electrical field with the spatially variable effective ``Hund's rule'' magnetic field yields
a charge density wave whose amplitude is balanced by the Coulomb interaction.  Similar noncollinear orderings
may exist in other systems like dilute magnetic semiconductors, where simple estimates predict a much stronger
version of the effect we find.  We show that the same type of nonequilibrium charge ordering is generic to
noncollinear magnets that support effective magnetic fields (from exchange interactions) that are inconsistent
with the current-free Maxwell equations.

As in the two-dimensional quantized Hall effect, there are finite transverse conductances in three-dimensional
transport that do not depend on relaxation and hence are ``intrinsic.''  Here we focus on the Karplus-Luttinger
anomalous velocity in magnetic materials with broken time-reversal symmetry (``$T$-breaking materials''), which
has been argued to explain the anomalous Hall effect in ferromagnetic semiconductors~\cite{niu2}.   A similar
intrinsic transverse conductance underlies the ``spin Hall effect'' predicted for some doped semiconductors
without broken time-reversal symmetry~\cite{zhang1,niu1,zhang2}. Current in all these situations flows
perpendicular to electric field and another vector: in the ordinary Hall effect, the other vector is magnetic
field; in the anomalous Hall effect, it is the local magnetic moment; and in the proposed intrinsic spin Hall
effect, it is the spin polarization that determines the flow of the current. Because the direction of the
current is determined by this third-party vector, it is interesting to ask what will happen if this vector is
not homogenous.

The standard derivation of the anomalous Hall effect contribution from the Karplus-Luttinger term
depends on nearly constant local moments, so the nearly ferromagnetic noncollinear magnets like
$MnSi$\cite{mnsi1,mnsi2} are natural candidates.  An intuitive understanding of the induced CDW we find in such materials can be gained by considering a stacking of several blocks of an itinerant ferromagnet (e.g., iron), in such a way that the local ferromagnetic direction rotates in a helix (Fig. 1).  Then under an applied electric field perpendicular to the helix direction, the different Hall currents in each block will lead to a buildup of charge at the junctions between different blocks.  The result we find for itinerant helimagnets can be viewed as a continuum version of the above physics, in which the transverse current flow and charge buildup is dominated by the anomalous Hall effect in materials with broken time-reversal and inversion symmetries.  This charge buildup will be largely compensated by screening by the conduction electrons, but there is a residual lattice distortion and charge oscillation in the helix direction that should be experimentally detectable.

Long-period modulations (periodicity over several hundred
angstroms) are also found in $SrFeO_3$ and $CaFeO_3$\cite{fe1}. We find that in these helimagnets and some other
configurations, an external electric field in certain directions will induce a charge density wave, and that
this nonequilibrium charge ordering is directly related to the symmetry of the magnetization. The estimates we
give of the strength of this charge density wave are easily carried over to other materials; as the anomalous
velocity only depends on the electrons or holes close to the Fermi surface~\cite{haldane}, the detailed band
structure is irrelevant.

Since we will consider materials with no symmetry under spatial inversion and with nonzero spin-orbit
coupling, the leading spin-orbit coupling should be linear in electron momentum. The free-electron Hamiltonian is
 \beq H = \sum_{{\bf k},\sigma}
c^\dagger_{{\bf k} \sigma} (\epsilon_{{\bf k}} + \gamma {\bf k } \cdot {\bf \sigma}) c_{{\bf k} \sigma}.
\label{hamiltonian}\eeq For simplicity, we will later approximate the dispersion relation by $\epsilon_{{\bf k}}
= \hbar^2/(2m_e)\cdot k^2$. The detailed structure of the spin-orbit coupling term is not too important for the
following calculation, and in fact the same effect appears for the Luttinger Hamiltonian for a spin-3/2 band
with only $({\bf k} \cdot {\bf s})^2$ terms.  The modified semiclassical equation of motion is \beqn
\hbar\dot{\bf k} = -eE-e\dot{\bf x}\times \bf B\cr\dot{\bf x} = \frac{\partial\epsilon_{\bf k}}{\partial \bf
k}-\dot{\bf k}\times\bf \Omega\label{motion}.\eeqn The Berry curvature $\Omega$ can be regarded as a magnetic
field in momentum space: in terms of the periodic part $u_n$ of the wavefunction for Bloch band $n$, its form is
\beq \Omega_i(n,k) = \epsilon_{ijk} {\rm Im} \left< {\partial u_n \over \partial k_y} \big| {\partial u_n \over
\partial k_x} \right>. \label{berry}\eeq We can easily see in (\ref{hamiltonian}) that the helicity is a
conserved quantity, and for the same momentum, electrons with different helicities have different energy, so
actually here we have two bands, with two different helicities. These two bands have a degenerate point at
$\mathbf{k} = 0$, close to this point, using (\ref{berry}), we will see the Berry phase looks like a monopole in
momentum space, \beq\Omega_i(\lambda,k) = \lambda\frac{k_i}{k^3}.\eeq Here $\lambda$ is the helicity ${\bf \hat
k} \cdot {\bf \sigma}$. Although we use this form for the Berry phase in our calculations,  the specific
form of the Berry phase is not important: the anomalous Hall effect is basically a Fermi surface effect. If there is a gap between two bands,  an infrared cut-off at $k = 0$ is necessary, but this will not contribute to the final result.

\momit{In this situation, detailed calculation tells us the magnetic field is very similar to
monopole \cite{zhang1}\beq\Omega(\lambda)_{k} = \lambda(2\lambda^2-\frac{7}{2})\frac{k_k}{k^3}.\eeq $\lambda$ is
the helicity, a conserved quantity in this spin-orbit coupling Hamiltonian.}

Under an electric field ${\bf E}$, the solution of (\ref{motion}) has both normal acceleration along electric
field and an anomalous velocity in the plane normal to ${\bf E}$, leading to an anomalous Hall current
\cite{niu2} \beq {\bf J} = \sum_{\lambda}\frac{e^2}{\hbar}\int d^3k ({\bf \Omega}(\lambda) \times {\bf E})
f(\varepsilon). \label{integral}\eeq Here $f(\varepsilon)$ is the distribution of particles.  If
$f(\varepsilon)$ is just the equilibrium distribution, the anomalous current is zero because the equilibrium
distribution is even under ${\bf k} \rightarrow -{\bf k}$, and the Berry phase is odd.  More generally, the
anomalous Hall effect is forbidden by time-reversal symmetry: electric field is even under time reversal, but
current is odd, and because the current is perpendicular to electric field, this anomalous Hall current has no
dissipation, so without another vector which is odd under time-reversal there can be no Hall current.  In
magnetic materials the ordered moments provide this additional vector, analogous to an external applied field.

The ordered moments produce the Hall current via the ``local Hund's rule'' coupling to spin, whose origin is in
exchange interactions rather than magnetic dipolar interactions.  This interaction between itinerant electrons
or holes and the local magnetic order can be modelled in a simple Hamiltonian \beq H^\prime = h{\bf m}\cdot {\bf
s}\label{spincouple}.\eeq Here ${\bf m}$ is the local magnetic moment. It is assumed that the background moments
are strongly ordered so that it is appropriate to use the expectation value of ${\bf m}$ as the Hund's rule
field and ignore fluctuations.  We will assume for now that the magnetic configuration has constant magnitude
and normalize $|{\bf m}|^2 = 1$.

The conduction electron spin density ${\bf s}$ includes spin-1/2 matrices in MnSi and spin-3/2 for the Luttinger
model of GaAs (here we define ${\bf s}$ to be dimensionless so that $h$ has units of energy). The exchange field
$h$ is proportional to the magnitude of the ordered local moment. If Mn spins are fully polarized, $h =
N_{Mn}SJ\label{exchange}$ \cite{niu2}, where $N_{Mn}$ is the density of Mn ions, $S$ is the spin of Mn ion,
which is 5/2, and $J$ is the exchange energy between local moment and itinerant electrons, which we take to be
$50\ {\rm meV\ nm}^3$\cite{exchange}.  In MnSi, the Mn ions are not fully polarized and the local moment is
about $0.4\mu_B$ per ion \cite{mnsi3}. So in this material $h$ is suppressed to 8\% of its maximum value.  In
MnSi, the lattice constant is 4.6 angstroms and there are 4 Mn ions per unit cell, so we obtain $N_{Mn} \simeq
30/nm^3$. Collecting all the results above, we finally estimate $h = 300\ {\rm meV}$ in MnSi.

On symmetry grounds, if both time-reversal and inversion symmetries are broken, one expects a Hall current from
the anomalous velocity term: \beq \mathbf{J}_{AH} = \sigma_{AH} \mathbf{E}\times\mathbf{m}\label{ahall}.\eeq
Both sides of this equation are odd under time reversal, so a nonzero anomalous Hall effect is consistent with
broken time-reversal and inversion symmetries.

In this problem we have three different energy scales: the Fermi energy, spin-orbit coupling energy, and the
Hund's rule coupling between itinerant spin and local moment. For convenience we assume $E_f\gg\lambda\gamma
k_f\gg\lambda h$, although our main result does not depend on this assumption. In this case the basic band
structure of this material is still applicable and we can regard the effect of local moment as a perturbation.
Suppose that ${\bf m}$ is along the ${\hat z}$ direction.  Then the Fermi sea with positive helicity will be
pushed down along $-\hat{z}$, because if an electron with momentum along negative $\hat{z}$ has positive
helicity, it will benefit from the interaction with local moment.  For the same reason, the Fermi sea with
negative helicity will be pushed up along $\hat{z}$. Thus the Fermi surface has an anisotropic Fermi wavenumber
$k_F$.

Because the energy at the Fermi surface must be constant, we have \beq \frac{\hbar^2}{2m_e}k^2_f +
\lambda\frac{\gamma}{2}k_f = \frac{\hbar^2}{2m_e}k_\theta^2 + \lambda\frac{\gamma}{2}k_\theta +
\lambda\frac{h}{2}\cos\theta.\eeq Solving these equations, we obtain \beq k_\theta = k_f -
\frac{hm_{e}\lambda\cos\theta}{\hbar^2 k_f}\label{fvector}.\eeq We have used the assumption that the Fermi
energy is much greater than the spin-orbit coupling to ignore the $\gamma$ in the denominator. So the unknown
value of $\gamma$ is irrelevant as long as it satisfies the assumption above.  This is analogous to the
situation in the Rashba model, which in the absence of interactions and disorder has a universal spin-Hall conductivity independent of spin-orbit coupling \cite{niu1}. Here $k_f$ is the Fermi vector at $\theta = \pi/2$, $m_e$ is the electron
effective mass, $\theta$ is the angle between $\bf k$ and $\bf m$, and $\lambda$ is the helicity. Now because of
this anisotropy, integrating over the Fermi sea as in (\ref{integral}) and summing over the two helicities
finally obtains a nonzero Hall current: the transverse conductivity is \beq\sigma_{AH} \simeq
\frac{e^2hm_e}{\hbar^3}\frac{1}{k_f}\label{ahall2}.\eeq For a numerical estimate, take $1/k_f \approx 1$ \AA\
and replace the effective mass by the bare mass: then the Hall conductivity $\sigma_{AH} \simeq 2\times 10^3
\Omega^{-1}cm^{-1}$. In very clean samples of MnSi, the diagonal conductivity $\sigma$ is about $10\ \mu
\Omega^{-1}$ cm$^{-1}$ below 20 K, the temperature below which the local moments exist\cite{mnsi1}, so
$\sigma_{AH}/\sigma\approx 10^{-4}$.  As discussed below, this ratio will increase in slightly disordered
samples.

This $1/k_f$ behavior appears strange because it predicts that as the particle density vanishes, the Hall effect
will be infinitely large. However, the above calculation is based on the assumption that the coupling with local
moment is much weaker than the Fermi energy. If $k_f$ is very small, the Hund's rule coupling dominates, almost
all the spins are aligned along the direction of local moment, and it is the spin-orbit coupling that is a
perturbation.

\momit{However, in this case, we don't have the original band structure any more, because all the spins tend to have
$S_z = 1/2$, and we only have one band. So the $1/k_f$ in (\ref{ahall2}) should not bother us because it is only
applicable when the fermi sea is big enough.}

All the calculations above are based on the nonvanishing Berry phase, which depends on adiabaticity of electron
motion.  Heuristically, because $\Delta t\cdot\Delta E\geq \hbar/2$, the relaxation time $\Delta t$ should be
large in order to make sure $\Delta E$ is very small compared with the band gap at Fermi surface. Although the
precise strength of spin-orbit coupling in the materials we study is unknown, it has to be much bigger than $h$,
the coupling to local moment, for the perturbative calculation above to be valid. So if $\tau \gg \hbar/h
\approx 10^{-14}s$, the calculation should be valid: this is a typical relaxation time in a metal in room
temperature, so at low temperature our assumptions are satisfied.

Now consider what will happen if the local moment is not perfectly collinear. First we discuss the helical case
which is relevant to MnSi and other materials discussed in the introduction\cite{mnsi1} \cite{fe1}. For
simplicity, we take local moments  \beq\mathbf{m}(x) = \cos(qx)\hat{z}+\sin(qx)\hat{y}\label{spiral}.\eeq The
modulation period $2\pi/q$ is typically at least 100 \AA, which means the local moment is rotating very slowly
when we move along $\hat{x}$. Because it is rotating very slowly, particles in different regions of the material
only see a homogenous local moment, and locally have the same momentum distribution as previous several
paragraphs. First keep the electric field along $\hat{y}$: then based on (\ref{ahall}) and (\ref{ahall2}), the
local Hall current density is \beq J_{AH}(x) = \cos(qx)\sigma_{AH}E_y.\eeq However, electric charge is always a
conserved quantity, and obviously, this anomalous charge current does not satisfy the static continuity
condition $\bf \nabla\cdot \mathbf{J}_{AH} = 0$, so charge will accumulate, inducing an electric field. Finally,
given some dissipation mechanism, the system will reach a steady state in which the diagonal current from the
induced electric field cancels the original Hall current: \beq\sigma E^\prime = -J_{AH}\label{CDW1}.\eeq Here
$\sigma$ is the normal diagonal conductivity of the material.  From the Maxwell equation $\mathbf{\nabla}\cdot
\mathbf{D} = 4\pi\rho,$ we obtain the relation between accumulation and electric field \beq \rho(x) =
q\sin(qx)\frac{4\pi\epsilon(q)\sigma_{AH}}{\sigma}E_y\label{cdw2}.\eeq $\epsilon(q)$ is the static dielectric
function at wave vector $q$.  This $\rho(x)$ describes the {\bf bare} or external electric charge density, which
will be strongly screened if $\epsilon(q) \gg 1$.

Let us estimate the magnitude of this charge density wave effect. First, in metal $\epsilon(q)$ is given by
$\epsilon(q) = 1+q^2_{TF}/q^2$, if $q^{-1} = 10\ nm$, we get $\epsilon(q) \simeq 2\times 10^4$. Also we have
$\sigma_{AH}/\sigma \simeq 10^{-4}$. If we take electric field to be $100$ V cm$^{-1}$, the density modulation
is about $10^{14}cm^{-3}$, which because of its periodicity should be measureable in a diffraction experiment.
Furthermore, from (\ref{cdw2}) the density modulation is proportional to the ratio between anomalous Hall
conductivity and diagonal conductivity.  Reducing the diagonal conductivity by adding impurities will lead to
increased charge modulation until the relaxation time becomes so short that the Berry's phase assumption
discussed above is violated.

\begin{figure}
\includegraphics[width=2.5in]{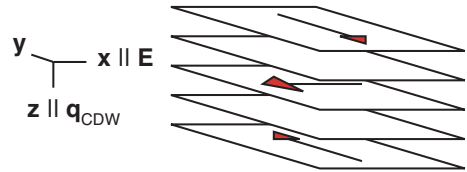}
\caption{Geometry of charge accumulation in a helical magnetic background.  In an electric field ${\bf E}$ perpendicular to the helix direction, a transverse CDW forms with the same modulation vector ${\bf q}$ as
the helix.} \label{figone}
\end{figure}

This charge density wave exists for helical direction $\hat{x}$ when the electric field has in any component in
the $yz$ plane. However, if the field is along $\hat{x}$, the Hall current will be in $yz$ plane, and it is
homogenous in this plane, so $\nabla\cdot \bf J = 0$ and there is no accumulation.  Actually, the formation of
charge density wave is because of the breaking of an approximate spiral symmetry possessed by the long-period
helical spin state.  A homogeneous background has 3 translational symmetries, $T_x$, $T_y$ and $T_z$.  When the
helical magnetic moment is added, it seems that $T_x$ has been broken. However, the system still has a special
helical or ``screw'' symmetry: combine the translation along $\hat{x}$ with a rotation, by defining \beq
T^\prime_x = T_x(a)R_x(aq),\eeq where $a$ is the lattice constant and $R_x(\theta)$ is the rotation
transformation around axis $\hat{x}$. We can see the system is invariant under $T^\prime_x$, $T_y$ and $T_z$,
and these three operators all commute. When we turn on the electric field in $yz$ plane, this spiral translation
symmetry is broken by the external field, and we have a charge density modulation which is forbidden if the
helical symmetry is unbroken., and the period is the same with the spiral moment configuration. However, if the
electric field is along $\hat{x}$, this spiral symmetry is not broken as long as the external field is
homogenous.  We can conclude that, although our calculation above was based on an assumption about the three
energy scales in the problem, this CDW effect should be relatively independent of these assumptions, because it
is controlled by symmetry breaking.

This absence of accumulation in the absence of an applied electric field is strictly correct in the continuum
limit where the helical transformations are exact symmetries: lattice effects can lead to a very small
accumulation if the helix vector is commensurate with the periodicity of the lattice.  In the continuum, charge
accumulation is always weak as long as the magnitude of the magnetic ordering is constant, because of a
local symmetry.


This absence of accumulation in the absence of an applied electric field is strictly correct in the continuum
limit where the helical transformations are exact symmetries: lattice effects can lead to a very small
accumulation if the helix vector is commensurate with the periodicity of the lattice.  In the continuum, charge
accumulation is always weak as long as the magnitude of the magnetic ordering is constant, because of a
local symmetry.  First
write (\ref{hamiltonian}) and (\ref{spincouple}) in operator form: \beqn H = \psi^\dagger_\alpha(-
\frac{\hbar^2}{2m_e}\nabla^2 \delta_{\alpha\beta}-\gamma(i\nabla\cdot \bf s)_{\alpha\beta})\psi_\beta \cr+h\bf
m(\vec{r})\cdot(\psi^\dagger_\alpha\bf s_{\alpha\beta}\psi_{\beta}).\eeqn Now in this Hamiltonian let the local
magnetization $\bf m$ vary slowly from point to point. The direction of local moments in the whole material can
be made uniform by performing a position-dependent rotation, written as \beq R(\vec{r})_{ij}m(\vec{r})_{j} =
m(0)_i.\eeq We can keep the interaction $H^\prime$ invariant if we perform another position-dependent $SU(2)$
transformation $U(\vec{r})$ on the spinor $\psi_\alpha$, as long as \beq U(\vec{r})^\dagger S_iU(\vec{r}) =
R(\vec{r})_{ij}S_j.\eeq Normally, this kind of symmetry is broken by the kinetic term because the derivative
will be modified to $\partial_i \rightarrow \partial_i-U^\dagger\partial_iU$ after a position
dependent transformation.

However, because we assumed the variation of local direction is slow, both $R_{ij}$ and $U$ are very slowly
varying functions, so we can ignore $U^\dagger\partial_iU$ in the kinetic term. This is equivalent to ignoring
derivatives of the $SU(2)$ transformation. Because the AHE only involves particles close to the Fermi surface,
as long as the characteristic length scale of variation of local moment is much larger than $1/k_f$, the
derivatives $U^\dagger\partial_iU$ do not modify the spectrum close to the Fermi surface and can be neglected.

Because of this local $SU(2)$ symmetry, the slowly varying system is equivalent to the case with homogenous
magnetization, and without electric field the charges are homogenously distributed. However, when the magnitude
of local moment varies, the $SU(2)$ gauge symmetry is broken as we can never perform any position dependent
rotation to transform this system to the homogenous magnetization case. When we add certain external fields to
the system with constant $|{\bf m}|$, the gauge symmetry is also broken: local rotation transforms the external
field, and without protection by symmetry, there will generically be an inhomogenous charge distribution.

In the helical case, an electric field along the helix direction preserves the symmetry and the charge density
remains homogenous. It is interesting to look for magnetic moment configurations in which the charges remain
homogenous for any applied electric field.  The constant magnetic moment is certainly an example, but there are
others: to prevent charge accumulation, the Hall current must satisfy $\nabla \cdot {\bf J}_{AH} = 0$.  This
gives a constraint on the moment configuration: \beqn \nabla\cdot \mathbf{J}_{AH} &=& \sigma_{AH} \nabla
\cdot(\mathbf{m}\times\mathbf{E}) \cr &=& \sigma_{AH}(\mathbf{E}\cdot\nabla\times {\bf m} -
\mathbf{m}\cdot\nabla\times\mathbf{E}).\eeqn The second term in the bracket of the second line is zero because
curl of electric field is always zero.  Hence there is no charge accumulation in any electric field if ${\bf m}$
has zero curl.  This requirement is satisfied if ${\bf m} = \nabla\phi$, so the moment configurations with no
accumulation are equivalent to source-free {\it electric} field configurations of constant magnitude.  One
example is the monopole configuration, in which $\mathbf{m}(\mathbf{r}) = \hat{r}$ with the potential $\phi =
|r|$. For this monopole background, the charge distribution remains homogenous for any applied field.

We conclude that the coupling between itinerant electrons and local moment gives rise to an instability in the
charge distribution in the presence of an applied electric field.  The simplest steady state of this
nonequilibrium problem has an inhomogenous charge density that, for a helical magnet, is a charge density wave
of the same period as the helix.  In general, magnetic configurations with constant magnitude but slowly varying
direction give rise to charge accumulation only in an electric field, while configurations with variable
magnitude cause charge accumulation in equilibrium.  However, there are special configurations where charge
homogeneity is protected for any an applied field.


\bibliography{ahall4}

\end{document}